\documentclass{emulateapj}

\shorttitle{High CO(3--2)/CO(1--0) ratio gas in NGC 604}
\shortauthors{Tosaki et al.}

\begin{document}


\title{Arc-like distribution of high CO($J$=3--2)/CO($J$=1--0) ratio gas 
surrounding the central star cluster of the supergiant H{\sc ii} region NGC 604}


\author{T.Tosaki\altaffilmark{1}, R. Miura\altaffilmark{2,3}, 
T. Sawada\altaffilmark{1}, N. Kuno\altaffilmark{1}, K. Nakanishi\altaffilmark{1},
K. Kohno\altaffilmark{4}, S.K. Okumura\altaffilmark{1} and 
R. Kawabe\altaffilmark{3}}




\altaffiltext{1}{Nobeyama Radio Observatory, Minamimaki, Minamisaku, Nagano, 384-1805, Japan}
\altaffiltext{2}{Department of Astronomy, The University of Tokyo, Hongo, Bunkyo-ku, Tokyo, 113-0033, Japan}
\altaffiltext{3}{National Astronomical Observatory of Japan, 2-21-1 Osawa, Mitaka, Tokyo 181-8588, Japan}
\altaffiltext{4}{Institute of Astronomy, The University of Tokyo, 2-21-1 Osawa, Mitaka, Tokyo, 181-0015, Japan}


\begin{abstract}
We report the discovery of a high CO($J$=3--2)/CO($J$=1--0) ratio gas 
with an arc-like distribution 
(``high-ratio gas arc'') surrounding the central star cluster of 
the supergiant H{\sc ii} region NGC 604 in the nearby spiral galaxy M 33, 
based on multi-$J$ CO observations of a 5$^\prime$ $\times$ 5$^\prime$ region 
of NGC 604 conducted using the ASTE 10-m and NRO 45-m telescopes.
The discovered ``high-ratio gas arc'' extends 
to the south-east to north-west direction with a size of $\sim$ 200 pc.
The western part of the high-ratio gas arc closely coincides well 
with the shells of the H{\sc ii} regions traced by H$\alpha$ and radio continuum peaks. 
The CO($J$=3--2)/CO($J$=1--0) ratio, $R_{3-2/1-0}$, ranges between 0.3 and 1.2 
in the observed region, and the $R_{3-2/1-0}$ values of the high-ratio gas arc are 
around or higher than unity, indicating very warm 
($T_{\rm kin} \geq$ 60 K) and dense ($n_{{\rm H}_2} \geq$ 10$^{3-4} {\rm cm}^{-3}$) 
conditions of the high-ratio gas arc.
We suggest that the dense gas formation and second-generation star formation 
occur in the surrounding gas compressed by the stellar wind and/or supernova 
of the first-generation stars of NGC 604, i.e., the central star cluster of NGC 604.
\end{abstract}


\section{Introduction}

Giant or supergiant H{\sc ii} regions (hereafter referred to as GHRs) are 
one of the most prominent objects
in star-forming galaxies at the optical wavelength.
Their sizes often reach a scale of a few 100 pc,
and their H$\alpha$ luminosities are typically of the order of 
a few $10^{39-40}$ erg s$^{-1}$, 
which corresponds to a few 10 to a few 100 O5 stars \citep{Ken84}.
The structure of GHRs is characterized by 
(1) the presence of a central young star cluster
and (2) extended shells and/or arc-like filaments surrounding the
central star cluster. 
For example, 30 Dor in Large Magellanic Cloud (LMC), 
which is the most luminous H{\sc ii} region in the Local group, 
houses the compact cluster R136, 
which is classified as a super star cluster (SSC) 
\citep{Hunt95}.
On the other hand, NGC 604 resides in M 33, 
which is the second-most luminous supergiant H{\sc ii} region after 30 Dor, 
hosts a scaled OB association \citep{Hunt96, Maiz04}. 
These central star clusters appear to have formed during the initial stages 
of the formation of GHRs, and are expected to have a strong impact 
on their natal molecular clouds due to 
their strong UV radiation, stellar wind, and supernova explosion.
Therefore, GHRs provide us with an ideal environment 
to understand the clustered OB star formation process,
and their impact on the ambient interstellar medium (ISM).
These physical processes are also crucial in the evolution of starburst 
in galaxies.

In this study, we present the $^{12}$CO($J$=3--2) and $^{12}$CO($J$=1--0) 
observations
of NGC 604 in the nearest face-on spiral galaxy M 33
using the Atacama Submillimeter Telescope Experiment (ASTE) 10-m and 
the Nobeyama Radio Observatory (NRO) 45-m telescopes.   
The structure of NGC 604 is complicated and containing many shells, filaments, 
and arc-like structures; furthermore,  
there is a central star cluster surrounded by arc-like H{\sc ii} regions 
\citep{Gome00} and $\sim$ 200 massive OB stars \citep{Gonz00}.
Several radio components exist in the arc-like H{\sc ii} regions, 
which are photoionized from the inside by the obscured massive stars, 
embedded in the further extended halo \citep{Chur99}.
The atomic and continuum emission of NGC 604 in the far-infrared wavelengths 
is also bright
\citep{hig03, hip03}.
Its proximity ($D$ = 0.84 Mpc; Freedman et al. 1991) and the favorable
inclination angle of the host galaxy ($i$ = $52^\circ$; Corbelli \& Salucci 2000)
allow us to clarify
the detailed distributions of young stars and the associated 
molecular clouds within this extreme star forming region.

Our objectives in conducting these molecular line observations are
(1) to investigate the spatial variation of the physical properties of molecular gas
and
(2) to understand their relationship with the GHR formation processes.
Several researchers have previously reported on observations of M 33 
in CO($J$=1--0) emission,
which can be collisionally excited even in low-density molecular gases 
such as $n$(H$_2$) $\sim$ $10^2$ cm$^{-3}$, 
and a large molecular cloud complex associated with NGC 604 
has been detected \citep{Wils92, Via92, Eng03, Hey04}.
Wilson et al.\ (1997) observed
$^{12}$CO($J$=2--1), $^{13}$CO($J$=2--1), and $^{12}$CO($J$=3--2) emission lines,
probing the denser components of molecular gas ($n$(H$_2$) $>$ $10^{3-4}$ cm$^{-3}$),  
toward two points in the NGC 604 region;
they found that the kinetic temperatures of the molecular clouds associated with 
the H{\sc ii} regions were systematically higher than those not associated 
with the H{\sc ii} regions.
Nevertheless, no map has been created for these high-$J$ CO lines thus far, 
and therefore the spatial variation of the physical conditions of 
the molecular clouds and their relationship with the central star cluster and 
the surrounding H$\alpha$ shells/arc-like structures have not yet been
well understood. 

Our new observations conducted using the ASTE 10-m and NRO 45-m telescopes provide 
high sensitivity $^{12}$CO($J$=3--2) and $^{12}$CO($J$=1--0) images of 
a $5' \times 5'$ or 1.3 $\times$ 1.3 kpc region located in the center of NGC 604
with an effective spatial resolution of $25^{\prime\prime}$ or 100 pc.
Using these new images, we report the discovery of 
a high CO($J$=3--2)/CO($J$=1--0) ratio gas with an arc-like distribution 
(high-ratio gas arc) surrounding the central star cluster of NGC 604.

\section{Observations}

We conducted $^{12}$CO($J$=3--2) observations using the ASTE 10-m submillimeter 
telescope located in the Atacama desert, Chile 
\citep{Ezawa04}
from 2006 July to August.
This was a part of our ADIoS project, which is an extragalactic CO($J$=3--2) 
imaging survey to obtain a global view of the dense molecular medium in galaxies 
\citep{koh07}.
We also conducted $^{12}$CO($J$=1--0) observations 
from 2005 December to 2006 March using the NRO 45-m telescope 
equipped with a $5 \times 5$ pixel focal-plane SIS 
array receiver (BEARS) capable of simultaneously observing 25 positions 
in the sky 
\citep{Suna00}. 
Further information on the 45-m observations 
will be reported in detail
in the forthcoming paper (Miura et al. 2007, in preparation).

The front-end for the ASTE observations was a single-pixel cartridge-type 
350 GHz SIS receiver (SC345; Kohno 2005). 
An XF-type digital spectrometer was used to cover a velocity width of 
445 km s$^{-1}$ with a velocity resolution of 5 km s$^{-1}$ at 345 GHz.
The observations were conducted remotely from the ASTE operation rooms
in National Astronomical Observatory of Japan (NAOJ)-Mitaka and NRO, Japan, using
the N-COSMOS3 network observation system developed by NAOJ \citep{Kam05}
\footnote
{Observations using the ASTE were conducted remotely from Japan
by using NTT's GEMnet2 and its partner R\&E (Research and Education) networks,
which are based on the AccessNova collaboration of University of Chile, 
NTT laboratories, and National Astronomical Observatory of Japan.}.
The typical system temperature in double-sideband (DSB) was 200 K. 
The absolute pointing accuracy and main beam efficiency were verified 
by observing the CO($J$=3--2) emission of o-Cet at interval of every 2 h. 
They were better than 2$^{\prime\prime}$ r.m.s. and 0.6 
during the observation runs, respectively.
The stability of efficiency also was monitored   
using o-Cet, and was found to be stable within $\pm 10\%$.

The on-the-fly (OTF) mapping technique was employed to obtain
the $^{12}{\rm CO}(J=$3--2) and $^{12}{\rm CO}(J=$1--0) data.   
The ``scanning noise'' was removed by combining the scan 
using the {\it PLAIT} algorithm described by \citet{Emer88}.

The full-width at half-power beams (FWHP) for the ASTE 10-m and NRO 45-m observations 
were 22$^{\prime\prime}$ and 16$^{\prime\prime}$ 
at the rest frequencies of $^{12}$CO($J$=3--2) (345 GHz) 
and $^{12}$CO($J$=1--0) (115 GHz), respectively.
We convolved these maps to a common spatial resolution of 25$^{\prime\prime}$;
this enabled us to measure CO ratios directly.

\begin{figure}
\epsscale{.50}
\plotone{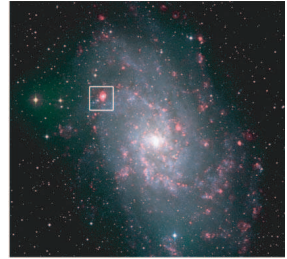}
\caption{
Three-color (B, V, and 6577\AA band (H$\alpha$)) composite image of M 33 
obtained using the 1.05-m Schmidt telescope installed at KISO observatory.
The image size is 59$^\prime$ $\times$ 55$^\prime$.
The white square represents the 5$^\prime$ $\times$ 5$^\prime$ area observed 
by the ASTE 10-m and NRO 45-m telescopes; this area shows the location of 
the giant H{\sc ii} region NGC 604.
\label{fig1}}
\end{figure}

\section{Results}

Figure 2 shows the velocity-integrated total intensity maps of 
$^{12}$CO($J$=1--0) and $^{12}$CO($J$=3--2) emissions,
integrated over a velocity width of 200 km s$^{-1}$ 
(from $V_{\rm LSR}$ = $-350$ km s$^{-1}$ to $-150$ km s$^{-1}$).
Although this velocity range was larger than that of the emission, 
$\sim$ 80 km s$^{-1}$, this has no effect on the total integrated intensity
due to the the baseline subtraction.

Wide-spread CO($J$=1--0) emission can be seen within
the mapped region. They are distributed around NGC 604, 
on the arm to its north, and on the downstream side of the arm.
The typical size of the clumps in the map is $\sim$ 100 pc, and
they aggregate to a larger complex that is similar to Giant Molecular Associations
\citep[GMAs;][]{Ran90}.
On applying the CLFIND method \citep{Will94} to the CO($J$=1--0) data
10 clumps were detected in this region and a few of them were new detections
not reported in previous CO($J$=1--0) observations.
Detailed properties of these CO clumps will be reported
in the forthcoming paper (Miura et al. 2007).

On the other hand, the $^{12}$CO($J$=3--2) map exhibits a compact morphology;
the major CO($J$=3--2) emission is confined to a small area close to 
the central cluster of NGC 604. 
Furthermore, the strong $^{12}$CO($J$=3--2) peaks are located to the north of 
the $^{12}$CO($J$=1--0) complex, i.e., the vicinity of the central star cluster 
of NGC 604.
Interestingly, no clear CO($J$=3--2) emission peak
can be observed near the second-strongest CO($J$=1--0) peak; 
this result reveals a significant variation in the physical properties 
of this region.

The molecular clouds around NGC 604 are located to the south of the center of 
NGC 604.
CO($J$=3--2) emission peak is closer to the central star cluster 
and surrounding the bright H{\sc ii} regions.
The southern side corresponds to the upstream side in NGC 604, 
taking the galactic rotation of M 33 into consideration.

We obtained the line ratio map of $^{12}$CO($J$=3--2) 
to $^{12}$CO($J$=1--0) emissions, $R_{3-2/1-0}$.
Figure 3 (left panel) shows the ratio of the integrated intensities of 
$^{12}$CO($J$=3--2) to $^{12}$CO($J$=1--0). 
The ratio was computed for the position 
where the signal to noise ratios of both lines exceed 2.
These ratios range from 0.3 to 1.3.
The maximum $R_{3-2/1-0}$ values in NGC 604 (1 -- 1.3) are similar to 
those in the center of the Milky Way \citep{Oka06} and 
the central region of M 83 \citep{Mura06}. 
On the other hand, the lower values are similar to 
those of the disk GMCs/GMAs in the Milky Way and the nearby spiral galaxy M 31 
\citep{Oka06, Tos06}.
In figure 3, we found an arc-like or shell-like distribution of a high-ratio gas 
surrounding the central star cluster of NGC 604.
This ``high-ratio gas arc'' extends to the south-east to north-west 
direction with a size of $\sim$ 200 pc.

\begin{figure}
\epsscale{0.8}
\plotone{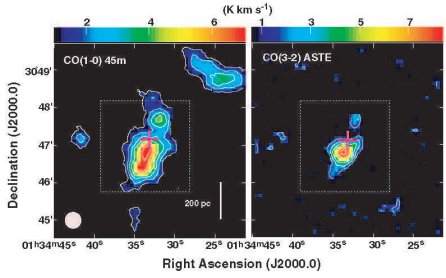}
\caption{
Velocity-integrated total intensity maps of 
the $^{12}$CO($J$=1--0) (left panel)
and $^{12}$CO($J$=3--2) (right panel) emissions
in the $5' \times 5'$ (1.3 $\times$ 1.3 kpc) region located in the center of 
NGC 604.
The rms noise levels, $\sigma$, in the CO($J$=1--0) and CO($J$=3--2), were 
0.5 and 1.3 K km s$^{-1}$, respectively.
The contour levels are 2, 4, 6, $\cdots$, and 14$\sigma$ for CO($J$=1--0),
and 1.5, 3.0, 4.5, $\cdots$, and 9$\sigma$ for CO($J$=3-2).
Note that the CO intensities are given in the main-beam temperature scale,
i.e., $\int T_{\rm mb}(v) dv$.
The dashed squares and crosses represent the region of the $R_{3-2/1-0}$ map 
shown in figure 3 and the position of the central star cluster, respectively.
These maps have a common spatial resolution of $25^{\prime\prime}$, indicated 
by a circle in the left-bottom corner.
\label{fig2}}
\end{figure}

\begin{figure}
\epsscale{.80}
\plotone{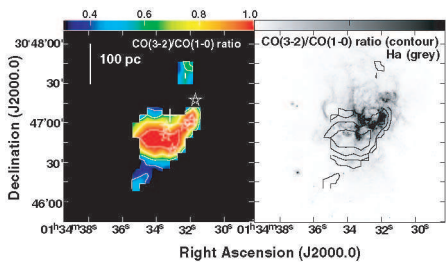}
\caption{
Maps of the ratio of $^{12}$CO($J$=3--2) to $^{12}$CO($J$=1--0) (left panel) 
and the H$\alpha$ emission by HST (right panel; from the HST archive). 
The contour levels are 0.4, 0.6, 0.8, and 1.0.
The crosshairs in the center indicates the position of the central
star cluster of NGC 604, while the stars represent the peak positions
of the H{\sc ii} regions detected by $\lambda$ 3.6-cm radio continuum emission
\citep{Chur99}. 
These maps clearly reveal the presence of high ratio gas 
with an arc-like morphology surrounding the central star cluster of NGC 604.
\label{fig3}}
\end{figure}

\section{Discussion}

\subsection{Nature of High-Ratio Gas Arc}

In this subsection, we discuss the physical conditions of 
the discovered high-ratio gas arc.

A model calculation using large velocity gradient approximation 
suggests that the high ratio gas
($R_{3-2/1-0} \sim 1$) requires a kinetic temperature of $T_{\rm kin} > 60$ K
and a gas density of $n$(H$_2$) $\sim$ $10^{3-4}$ cm$^{-3}$
(figure 9 in Muraoka et al. 2007).
These values are in very good agreement with the results 
of Wilson et al. (1997) for the cloud NGC 604-2.
These indicate that the warm and dense gases are distributed around 
the central star cluster with an arc-like morphology.

In addition, there exists an arc-like distribution of H$\alpha$ emission 
\citep{Gome00},
wherein several compact H{\sc ii} regions  
are embedded, as revealed by radio continuum observations \citep{Chur99}. 
These distributions exhibit a striking similarity to the western part of 
the high-ratio gas arc,
suggesting that massive stars are formed within the warm and dense gas layers
depicted as the high-ratio gas arc.

Here it should be noted that the high $R_{3-2/1-0}$ values in the arc
are due to not only the high kinetic temperature but also the high gas density.
If the high $R_{3-2/1-0}$ values are attributable solely to the UV heating 
from these newly formed stars,
the high line ratio can only coincide with the H$\alpha$ shells, 
i.e., the high-ratio gas 
should be observed predominantly in the western part of Fig. 3.
However, our line ratio map indicates the presence of high-ratio gas,  
even in the eastern part of the molecular clouds, which is set apart from 
the bright H$\alpha$ shell seen in Fig.~3 (right panel). 
Furthermore, another calculation conducted using a photo dissociation region (PDR) 
supports that the high $R_{3-2/1-0}$ values could be attributable to the high gas density
\citep{Kauf99}.

In conclusion, we suggest that the high-ratio gas arc reveals the presence of
warm and dense molecular gas layers, and the massive stars now form
within the dense gas arc.

\subsection{Triggered Dense Gas and Star Formation 
by ``First-Generation'' Star Formation}

In this subsection, we discuss the relationship among the high-ratio gas arc, 
the central star cluster, 
and the embedded young stars accompanying the H$\alpha$ shell.

The observational results obtained in this study 
can be explained by the following scenario.
First, stars were formed in the northern part of GMA; this is referred to 
as ``first-generation star formation''.
These stars can now be observed as the central star cluster.
Next, their stellar wind and supernova compressed 
the surrounding ISM. 
As a result, a dense gas layer was formed therein;  
such dense gases with high $R_{3-2/1-0}$ ratios are distributed around 
the central star cluster.
This is observed as the arc-like distribution of the high $R_{3-2/1-0}$ ratio gas.
New stars were then formed within the dense gases layer;
this is referred to as ``second-generation star formation'' 
triggered by the first-generation stars (i.e. the central star cluster).
These secondary young stars formed in the dense gas layer 
can now be observed as the H$\alpha$ shells and compact radio continuum sources 
as seen in Fig 3.
Figure 4 summarizes the schematic view of the proposed scenario for NGC 604.
A similar situation is also reported in the N 11 region, which is 
the second largest star forming region of LMC \citep{Hatano06}.

\begin{figure}
\epsscale{.70}
\plotone{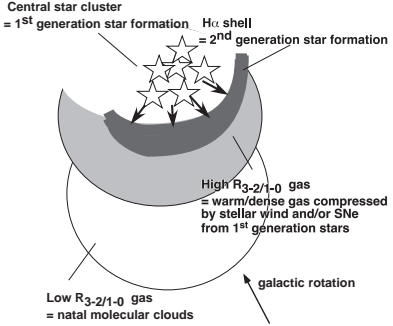}
\caption{Schematic view of NGC 604. 
\label{fig4}}
\end{figure}

Certain observations provide supporting evidence for the proposed scenario.
First, the velocity field derived from the H$\alpha$ emission of NGC 604 indicates 
that it consists of many filaments formed by
multiple blowout events 
due to the energy injected by massive stars \citep{Teno00}.
No H$\alpha$ emission or CO emission peaks are seen in the central star cluster; 
however, the arc-like structure of the high $R_{3-2/1-0}$ ratio gas has been 
associated with both H$\alpha$ and CO emissions.
These facts suggest that the H{\sc ii} regions embedded in the high-ratio gas arc 
are in fact on-going star-forming regions, 
while the central star cluster that has already dispersed ISM
is the ``past'' star-forming region.

Second, the presence of shocks in the H$\alpha$ shell 
was observed by optical spectroscopy 
\citep{Gome00}; 
these are attributed to the detected expanding motion \citep{Teno00}. 
This could be consistent with the view 
that the ISM surrounding the central star cluster 
was compressed by its energy blowout.
The observation of a compact H{\sc i} hole suggests the presence of SNR in NGC 604
\citep{deul90};
further, the observation of X-ray emitting hot plasma \citep{mis06} 
may also support such expanding motion.

Further evidence is provided by the optical narrow-band imaging 
of [S{\sc ii}] and [O{\sc iii}] emissions \citep{Teno00}.
In the arc-like region with the high ratio, the [O{\sc iii}]/H$\alpha$ ratio is high 
while the [S{\sc ii}]/H$\alpha$ ratio is relatively low.
Because both high [O{\sc iii}]/H$\alpha$ and low [S{\sc ii}]/H$\alpha$ ratios 
correspond to a high excitation condition,
it is suggested that the stars in the high-ratio gas arc region 
are significantly younger than those in other regions. 
This difference between the stellar ages of 
the central star cluster and the surrounding young stars  
is consistent with the proposed scenario.

On the other hand, the age of the central cluster of NGC 604 has been estimated to 
be $\sim 3 \times 10^6$ years \citep{Bruh03, Gonz00}.
Therefore, the timescale required to expand to the size of 
the detected H{\sc ii} region, as indicated 
by the radio continuum observations,
would be in the range of $10^6$ years to $8 \times 10^6$ years \citep{Chur99}.
We found no significant difference between the central star cluster ages and 
the radio-derived ages.

However, it should be noted that the stellar age was estimated over the region 
including both the high-ratio gas arc region and the central star cluster, 
i.e., the derived stellar age is a mixture of the proposed second-generation stars 
(in the H$\alpha$ shell) and the first-generation ones (i.e., the central cluster).
Consequently, inconsistency between these timescales and the proposed scenario is 
therefore not decisive 
and we need to conduct further observations to obtain spatially resolved stellar age 
measurements across the NGC 604 region.

Note that the compressed warm dense gas is seen only in 
the southern and western parts of the star cluster.
It is due to the direction from which material flows, namely, 
from the south-west to the north-east in spiral arm, 
and expanding material from first generation stars meets 
material flowing from the south-west into the arm. 
It suggests that a shock moves from the north-east to the south-west.

Thus, we can conclude that the arc-like structure of the high-ratio gas
is the site of second-generation star formation 
triggered by first-generation star formation.
NGC 604 is an example of a large-scale sequential star formation.

\acknowledgments

We would like to thank the staff of the Nobeyama Radio Observatory and 
the ASTE team
for their kind support in conducting our observations.
This study was financially supported by the MEXT Grant-in-Aid 
for Scientific Research on Priority Areas No.\ 15071202. 
The Nobeyama Radio Observatory is a branch of 
the National Astronomical Observatory of Japan, 
National Institutes of Natural Sciences (NINS). 
We also would like to thank Dr. Shingo Nishiura for providing us 
with the optical image of M 33 acquired using
the 1.05-m Schmidt telescope installed at the KISO observatory, 
operated by the Institute of Astronomy, University of Tokyo.



\end{document}